\documentclass[pre,amsfonts,amssymb,amsmath,reprint,showkeys,nofootinbib, twoside]{revtex4-1}
\usepackage[english]{babel}
\usepackage[utf8]{inputenc}
\usepackage{amsthm}
\usepackage{mathtools}
\usepackage{physics}
\usepackage{xcolor}
\usepackage{graphicx}
\usepackage[normalem]{ulem}
\usepackage[left=23mm,right=13mm,top=35mm,columnsep=15pt]{geometry} 

\usepackage{adjustbox}
\usepackage{placeins}
\usepackage[colorinlistoftodos, color=green!40, prependcaption]{todonotes}
\usepackage[pdftex, pdftitle={Article}, pdfauthor={Author}]{hyperref} 
\begin{document}

\title{Aerial mucosalivary droplet dispersal distributions with implications for disease mitigation}

\author{Brian Chang}
\email{brchang@clarku.edu}
\affiliation{Department of Physics, Clark University, Worcester, MA 01610, USA}
\author{Ram Sudhir Sharma}
\author{Trinh Huynh}
\author{Arshad Kudrolli}
\email{akudrolli@clarku.edu; Corresponding author}
\affiliation{Department of Physics, Clark University, Worcester, MA 01610, USA}
\date{Accepted December 8, 2020, Physical Review Research} 

\begin{abstract}
We investigate  mucosalivary dispersal and deposition on horizontal surfaces corresponding to human exhalations with physical experiments under still-air conditions. Synthetic fluorescence tagged sprays with size and speed distributions comparable to human sneezes are observed with high-speed imaging. We show that while some larger droplets follow parabolic trajectories, smaller droplets stay aloft for several seconds and settle slowly with speeds consistent with a buoyant cloud dynamics model. The net deposition distribution is observed to become correspondingly broader as the source height $H$ is increased, ranging from sitting at a table to standing upright. We find that the deposited mucosaliva decays exponentially in front of the source, after peaking at distance $x = 0.71$\,m when $H = 0.5$\,m, and $x = 0.56$\,m when $H=1.5$\,m, with standard deviations $\approx 0.5$\,m. Greater than 99\% of the mucosaliva is deposited within $x = 2$\,m, with faster landing times {\em further} from the source.  We then demonstrate that a standard nose and mouth mask reduces the mucosaliva dispersed by a factor of at least a hundred compared to the peaks recorded when unmasked.
\end{abstract}


\maketitle

\section{Introduction} \label{sec:introduction}
Dispersal of infected droplets during expiratory events such as coughing and sneezing is an important route for the transmission of tuberculosis, influenza, COVID-19, and other respiratory diseases~\cite{Bin2015,Wigginton2020}. According to the Centers for Disease Control and Prevention, coronavirus SARS-CoV-2, which causes COVID-19, is thought to spread when exhaled droplets from an infected person are inhaled into the lungs, or land on the faces of people who are nearby~\cite{CDC2020a}. A wide range of droplet sizes from submicrons to millimeters have been observed for the various exhalation modes with larger total volume and droplet sizes observed in the case of coughing and sneezing compared to breathing and talking~\cite{Chao2009,Tang2013,Anfinrud2020,Stadnytskyi2020}. 
Because the virus size is on the order of 100\,nm~\cite{Neuman2006,Yao2020}, even micron sized droplets can carry a significant number of viruses~\cite{Milton2013,Yan2018}. The exhalation speeds can vary greatly from 1.4 m/s for nasal breathing to 4.5 m/s while sneezing and coughing with droplet clouds observed to travel at least 0.6\,m~\cite{Chao2009,Tang2013}. 

It is generally accepted that a minimal amount of virus is required for a healthy person to be infected, but the actual numbers vary with disease. Reducing direct exposure, wearing masks, and avoiding poorly ventilated spaces are considered as further mitigation strategies in the spread of infectious disease~\cite{Bahl2020b,allen2020recognizing,chu2020physical,morawska2020can}.  Based on droplet dispersal ranges and infectious disease transmission studies going back over 80 years~\cite{wells1934air}, 2\,m or about 6 feet are given as a practical guide for prevention of transmission~\cite{CDC2020a}. However, small suspended droplets or aerosols can disperse further distances and the efficacy of this rule remains much debated in the context of COVID-19~\cite{morawska2020time,chagla2020airborne,Bartoszko2020,morawska2020reply} as it may underestimate the region of transmission 
in confined spaces~\cite{CDC2020c}. 

Despite the importance in determining prevention strategies, direct observation of droplet dispersal and surface deposition remain poorly characterized from a physical perspective. Beyond considering the difficulty of doing measurements with human subjects and large human to human differences, event to event variability even in an individual makes systematic investigations difficult. Analytical and numerical studies have been reported on droplet dispersal distances which consider drag~\cite{Cummins2020, Das2020}, but interactions between droplets mediated by the air were not considered. Such interactions are noteworthy because expiratory events like sneezing and talking cause turbulent puffs that create correlated motions~\cite{Dbouk2020,Abkarian2020}. Few systematic studies exist on how far and wide the actual mucosaliva can be transported by exhalations~\cite{Bourouiba2014,Anfinrud2020,Abkarian2020}. However, a difficulty is that the small aqueous droplets evaporate rapidly under typical atmospheric conditions, which makes their detection with direct high speed imaging also increasingly susceptible to systematic error. 

\begin{figure*}
\begin{center}
\includegraphics[width=.9\linewidth, trim={0 8.5cm 7cm 0},clip]{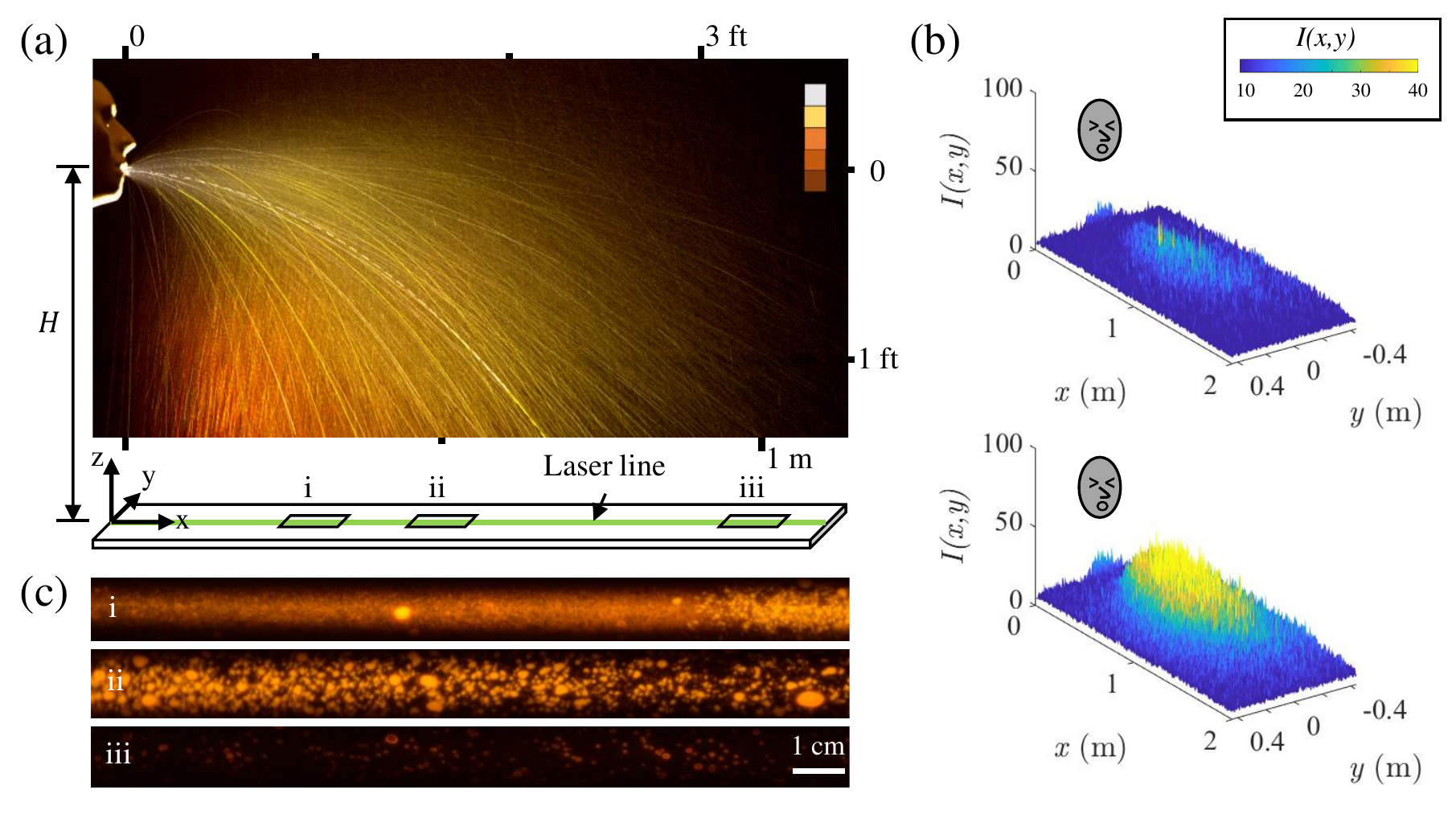}
\caption{ (a) Superimposed images of a mucosalivary spray emerging from the mouth of a 3D printed mannequin face obtained at 1000 fps over 4 seconds. Color bar indicates progression of time from white $t=0$\,s (white) to $t=4$\,s (orange). The mucosaliva is marked with a fluorescent tracer and deposits over time onto a horizontal surface. (b) The light intensity, $I(x,y)$, corresponding to mucosaliva deposited on the surface in front of the head after $n_s = 5$ (top) and $n_s =20$ (bottom) expirations, or sprays ($H = 0.5$\,m). Images are taken approximately 10 seconds after the last spray. A deposition lobe directed along the symmetry axis in front of the face can be observed. The laser line is set at $y=0$ \,m. (c) Images of deposited mucosalivary fluid after $n_s = 20$ sprays illuminated by a 1\,cm wide laser light sheet at distance from source of (i) $x = 0.35$\,m, (ii) $x=0.5$\,m, and (iii) $x=1.0$\,m.  }
\label{fig:1_schematic}
\end{center}
\end{figure*}

Here, we examine mucosalivary sprays generated by a mechanical device as they spread in the air and get deposited on a horizontal substrate. The sprays generated under well-defined laboratory conditions are demonstrated to show dynamics consistent with published results on human sneezes. By adding a fluorescent dye as a passive tracer contaminant {instead of the viruses, we perform sensitive measurements for their dispersal distance}.
We find that the amount of deposited mucosaliva in front of the source decreases exponentially after reaching its peak, and at least 99\% of the material is deposited on the substrate within 2\,m. From a height of 1.5\,m, 50\% of the material lands in about 2~seconds, the remaining continues to fall slowly over at least 6~seconds depending on the distance to the source. We show that a buoyant cloud model captures the speed with which a majority of the droplets settle. Building on these measurements, we quantify the effect of wearing a mask on mucosalivary dispersal. In particular, we demonstrate that wearing even an inexpensive non-medical mask, {besides a N95 mask}, leads to the reduction of mucosaliva deposited on surfaces by a factor of at least a hundred. 

\section{Experimental System}
We synthesize the mucosalivary fluid corresponding to a healthy adult by mixing mucin from bovine submaxillary glands (BSM; Sigma-Aldrich) with distilled water at a concentration of 150\,mg/mL for at least two hours~\cite{Kesimer2017}. The medium is also marked with Rhodamine B which acts as the passive tracer representing virus particles. Rhodamine B in an aqueous medium has a peak fluorescence at 568\,nm when illuminated by a 532\,nm laser. When imaged through a band pass filter, the fluorescent light passes through while the bright laser line is removed, enabling a far more sensitive imaging of the deposited material on a surface versus reflected light. Additionally, while the water content in a droplet can decrease due to evaporation, Rhodamine B {stains the mucin, which} is not observed to evaporate or sublimate and thus is a more accurate representation of the virus contained within infected mucosaliva than simply using water. {We note that the overall intensity of the droplet decreases less than 5\% after the water evaporates over 900\,s in a large $\approx2$\,mm droplet (see Supplemental Documentation (SD) Movie~1~\cite{sup-doc}.)}

{The mucosalivary sprays are generated by pulling the medium into a small air-tight chamber, and under a piston-line action, is pushed out through a nozzle to create droplets. A fixed volume, $V_s = 0.75$\,mL, is expelled out at a rate of approximately 3.75\,mL/s
}
through the mouth of a 3D printed face shown in Fig.~\ref{fig:1_schematic}(a). {By varying the aperture size of the nozzle, we can change the appearance of the spray from a fine mist to large droplets.} The sprays are launched at heights of $H=0.5$, 1.0, and 1.5 m above a horizontal surface, corresponding to the distance between a table to a sitting adult, table to a standing adult, and floor to a standing adult, respectively.
The event is imaged using a Phantom VEO-E 310 camera (see Appendix~\ref{sec:appendix}) and a typical example corresponding to $H = 1.5$\,m is also shown in Fig.~\ref{fig:1_schematic}(a). Here, all images within the first 4\,seconds are combined and colored according to their time stamp (white being early and dark orange being later) to illustrate the overall droplet dynamics. While the size, speed, and amount of spray can be varied in our apparatus, a combination of mucosalivary volume $V_s \approx 0.75$\,mL, spray speed $u_0 \approx 5$\,m/s, and duration $T_s \approx 200$\,ms was obtained by trial and error using high speed imaging until these values match human sneezes found in literature~\cite{Bourouiba2014,Bahl2020}.

\begin{figure*}
\begin{center}
\includegraphics[width=.9\linewidth, trim={0cm 1.5cm 0cm 0}]{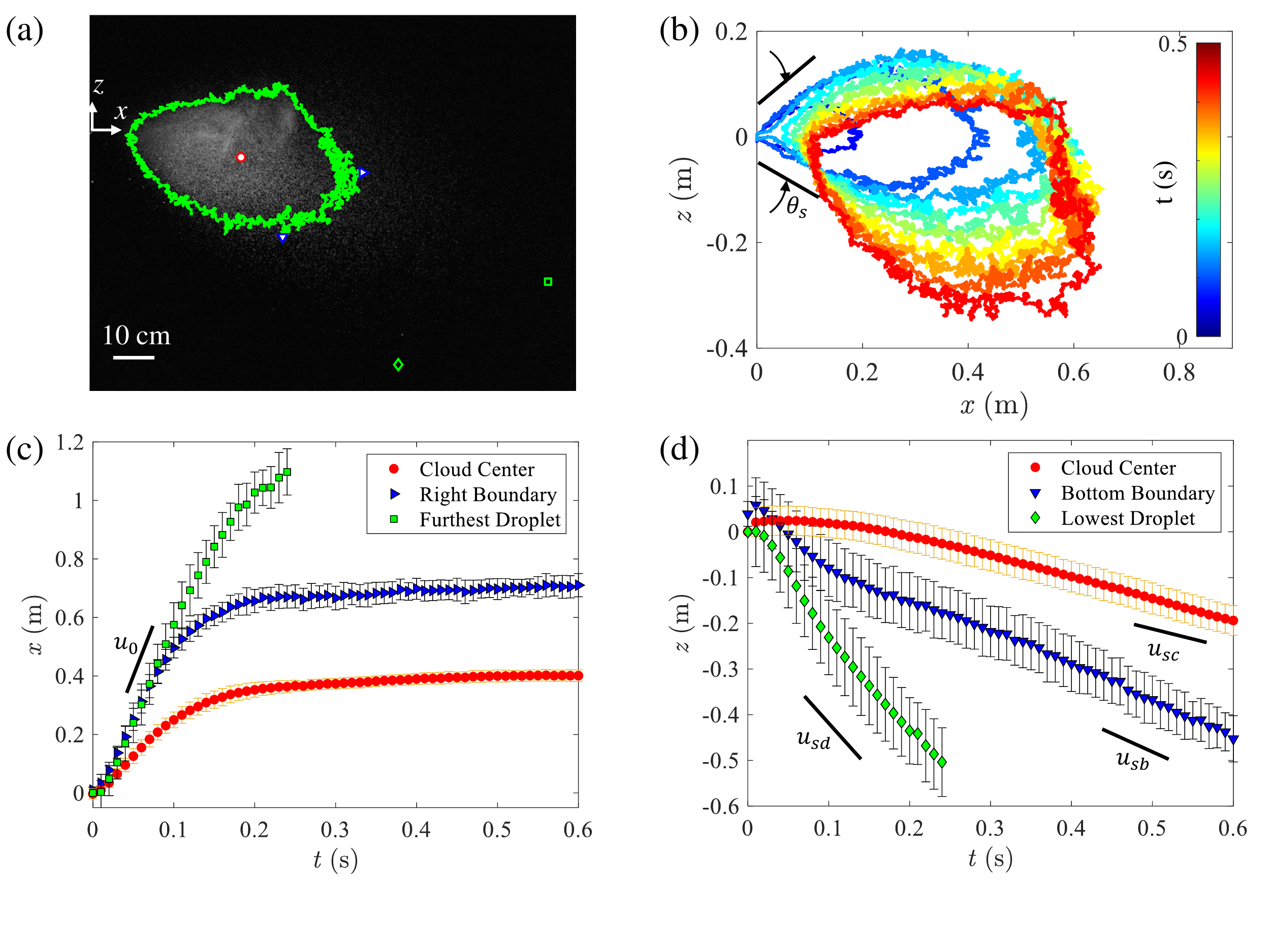}
\caption{(a) Side view snapshot of a mucosalivary spray at $t=0.35$\,s with a cloud threshold boundary (green line), where  
{\color{red} $\circ$} marks the cloud geometric center, and 
{\color{blue} $\triangleright$} and {\color{blue} $\triangledown$} 
mark the furthest $x$ extent and the lowest $z$ extent of the boundary, respectively. The green square and diamond indicates the furthest and lowest visible droplets in the $x$- and $z$-directions, respectively. (b) Tracked boundary of the cloud over time and the cone angle $\theta_s$. The geometric center of the cloud asymptotes towards $x=0.4$ m, whereas individual droplets separate from the cloud travel further. The initial propagation speed is $u_0=5$ m/s. (d) The settling speed for the center, boundary, and individual droplet is found to be $u_{sc} = 0.5$ m/s, $u_{sb} = 0.8$ m/s, and $u_{sd} = 2.0$ m/s, respectively. Each point represents the mean over $n_t = 10$ trials with error bars of one standard deviation.
}
\label{fig:2_cloudboundary}
\end{center}
\end{figure*}

The deposition of the mucosaliva on the horizontal surface is imaged with a camera which views a 2\,m long and 1\,m wide area in front of the face. The cumulative effect of multiple events can be examined by generating a number of sprays, $n_s$, at well defined intervals. This enables us to not only examine trace amounts of deposited material with greater sensitivity, but to also examine the effect of droplet interactions. We have found that the deposition is essentially independent of the time interval between sprays beyond one second, and use this minimum time interval in our investigations. Examples of deposited mucosaliva after $n_s = 5$ and $n_s = 20$ sprays are shown in Fig.~\ref{fig:1_schematic}(b). An elongated dispersal lobe is observed with greatest extension along the symmetry $x$-axis in front of the face. 

All experiments were performed in an enclosed room maintained at $23 \pm 2^\circ$C {with 65\%-75\% RH}, and while the HVAC system was turned off to investigate dispersal under well defined still-air conditions. Without these precautions, the lobe shape is found to be influenced by the direction of drafts. 

To investigate the dispersal with even greater sensitivity, a 1\,cm wide laser line (532\,nm, 40\,mW) is used to illuminate the mucosaliva deposited on the substrate directly in front of the head along the direction of the greatest range, as shown schematically in Fig.~\ref{fig:1_schematic}(a). Sample close up images obtained using a Nikon D3300 color camera at three different locations are shown  in Fig.~\ref{fig:1_schematic}(c).  Several discrete droplets of various sizes are observed to be deposited at the three different locations. We find that the mucosaliva droplet sizes range between $10 - 1000\,\mu$m, which is similar to those reported in a human sneeze~\cite{Han2013}. Further details on the calibration of the intensity to the volume of the medium and the various imaging methods used in the study can be found in Appendix~\ref{sec:appendix}. 

\section{Aerial Dynamics} \label{sec:cloud}
Figure~\ref{fig:2_cloudboundary}(a) shows a snapshot from the Supplemental Documentation (SD) Movie~2~\cite{sup-doc} of the mucosalivary spray obtained with the Phantom camera at time $t = 350$ ms with a 1\,ms exposure. The spray can be observed to emerge uniformly as a cone and becomes inhomogeneous as instabilities develop as the fast moving spray invades the quiescent air, before spreading as a cloud and settling over time. As is also clear from Fig.~\ref{fig:1_schematic}(a), it is apparent that a wide range of dynamics can be observed; droplets with fast, parabolic trajectories and droplets with slow, cloud-like motions. To quantify the observations, and compare these mechanically generated sprays to those reported in the case of human sneezes, we perform image processing to identify several features for characterization and comparison. Fig.~\ref{fig:2_cloudboundary}(a) shows an example of a boundary which encloses the multi-phase cloud identified by image processing implemented in MATLAB. This is accomplished by subtracting the background noise using images before the spray is initiated, converting the resulting gray scale image to a binary image under a threshold criteria, and then tracing the largest exterior boundary using the {\em bwboundaries} MATLAB function. The geometric center of the cloud projected onto the $x-z$ coordinate plane is calculated based on the outline boundaries, and its furthest extension along $x$ and $z$ are also detected and shown in Fig.~\ref{fig:2_cloudboundary}(a). 

A sequence of the identified cloud boundary over time is plotted in Fig.~\ref{fig:2_cloudboundary}(b). The cone angle $\theta_s$ of the emerging mucosalivary spray is marked in Fig.~\ref{fig:2_cloudboundary}(b) before the cloud detaches from the face at about $t = 200$\,ms. The spreading angle starts as $\theta_s= 50^\circ$ and grows to a maximum of $\theta_s= 70^\circ$. Both the cloud detachment time and initial spreading angles from the mechanically generated sneeze are similar to those found in published human sneeze movies~\cite{Bourouiba2014,Bahl2020,Gupta2009}. 

 However, it should be noted that this boundary is only an approximation to capture the region with a high density of droplets, and that several more isolated droplets can be observed outside the marked boundary in Fig.~\ref{fig:2_cloudboundary}(a). {We use an optimum threshold value that reduced an individual trial's variation over time. } The furthest droplet along the $x$-axis and $z$-axis are denoted on the image, and can be observed to be well ahead of the cloud in this time snapshot. 
We plot the furthest visible signature of droplets as a function of time $t$ alongside the propagation of the cloud 
along the $x$-axis and $z$-axis in Fig. \ref{fig:2_cloudboundary}(c) and Fig. \ref{fig:2_cloudboundary}(d), respectively. (While the cloud center and boundary is sensitive to the choice of threshold, we determined an optimal threshold value that yielded the least trial to trial variation.) These individual droplets isolated from the cloud can maintain their initial horizontal speed and travel further than the cloud counterpart. This two-component nature of the ejection is also noted for sneezes~\cite{Scharfman2016}. 

Based on the propagation speed of the furthest droplets and the cloud boundary shown in Fig.~\ref{fig:2_cloudboundary}(c), we find that the initial speed with which the droplets emerge is $u_0=5$\,m/s. This agrees well with human sneezes, where a sneeze may produce droplets traveling at an average velocity of 3.5\,m/s, and that 80$\%$ of droplets have velocities that are 5 m/s or lower~\cite{Bahl2020}. 

The evolution of the droplets' vertical location is shown in Fig.~\ref{fig:2_cloudboundary}(d). After the initial spray period of 200\,ms, the cloud's center reaches a settling velocity of $u_{sc}=0.5$\,m/s, whereas the lowest point of the boundary reaches a velocity of $u_{sb}=0.8$\,m/s. On the other hand, individual droplets, separate from the cloud, reach settling velocities up to $u_{sd}=2.2$\,m/s. Thus, clear differences between the fate of the cloud and the individual droplets occur over time. As a result, the cloud's center approaches a vertical asymptote around $x=0.4$\,m, whereas the furthest extent of the cloud boundary approaches $x=0.7$\,m. However, these measurements which focus on the cloud underestimate the actual mean dispersal distances as we see next with fluorescence imaging. This occurs because they systematically discount larger droplets as they fly beyond the cloud as well as very small droplets which do not scatter sufficient light and can also evaporate systematically faster when considering time periods over seconds~\cite{li2018modelling}. 

\section{Surface Deposition} \label{sec:distrubutions}
\subsection{Spatial Distribution}
\begin{figure*}
\begin{center}
\includegraphics[width=.95\linewidth, trim={0 6cm 0cm 0}]{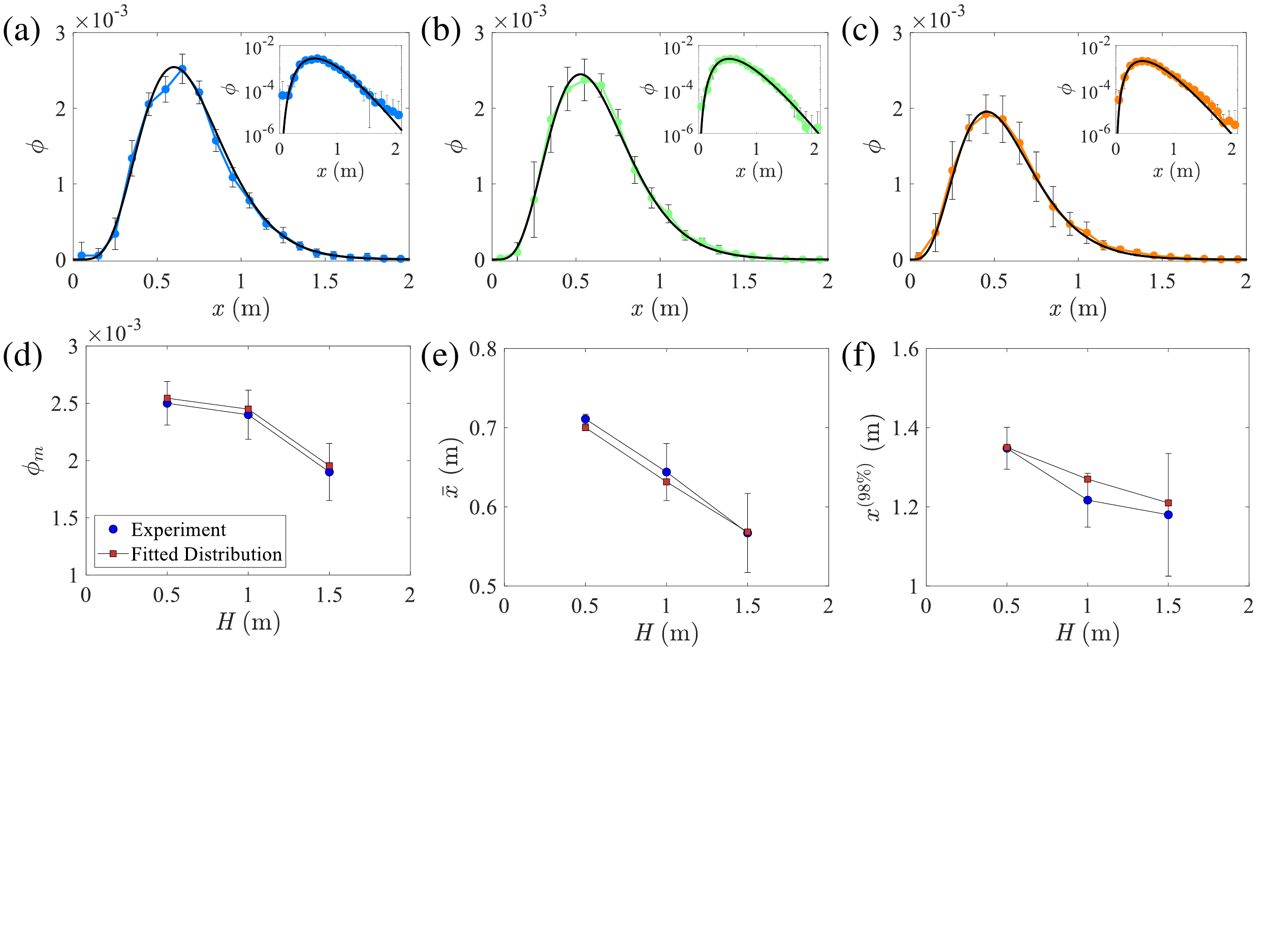}
\caption{Distribution profile of mucosalivary droplets deposited on a horizontal surface along $x$-axis from heights of (a) $H=0.5$\,m, (b) $H=1$\,m, and (c) $H=1.5$\,m ($n_s = 20$). Insets are the respective semi-log plots. The noise floor is $\phi_{noise}=4.4\times10^{-6}$, which is three orders of magnitude less that the maximum signal.
A total of $n_t=12$ trials are conducted for each height. Data points are binned with a bin size of 0.1 m. Distribution width $\sigma_x \approx 0.5$\,m for all heights. Black line is a fitted exponential function. (d) Maximum amount of material, $\phi_{m}$ decreases with height. (e) The mean of the distribution, $\bar{x}$, decreases linearly with increasing height. (f) Distance $x^{(98\%)}$ at which $98\%$ of the material is deposited decreases with $H$. Each point represents the mean over $n_t=12$ trials with error bars of one standard deviation.}
\label{fig:3_distributions}
\end{center}
\end{figure*}

The distributions of the deposited mucosaliva directly in front of the face using fluorescence imaging are shown for three different $H$ in Fig.~\ref{fig:3_distributions}(a-c). For each $H$, $n_t = 12$ trials with $n_s = 20$ sprays for each trial were performed to obtain the average distribution and the fluctuations from trial to trial. The volume of the droplets, $V$, along the $x$-axis is normalized by the total volume for 20 sprays, $V_{20} = n_s V_s \approx 15$\,mL. Therefore, the parameter $\phi = V/V_{20}$, is interpreted as a relative volume distribution, or a percentage, of mucosaliva deposited directly in front of the source along the $x$-axis. 
A broad distribution is observed in each case with a peak which systematically decreases as $H$ increases, showing a broader dispersal with expulsion height due to increased residence in air. As quantified further in Fig.~\ref{fig:3_distributions}(d), we find that the peak of the distribution, which we denote as $\phi_m$, decreases by 24\% (from $\phi_m=2.5\pm0.19 \times 10^{-3}$ to $\phi_m=1.9\pm0.25 \times 10^{-3}$) with this three fold variation in expulsion height $H$. Similarly, as shown in Fig.~\ref{fig:3_distributions}(e), the mean dispersal distance $\bar{x}$ decreases by 20\% from $\bar{x}=0.711\pm 0.006$\,m at $H=0.5$\,m to $\bar{x}=0.567 \pm 0.050$ m at $H=1.5$\,m. Fig.~\ref{fig:3_distributions}(e). 
Whereas, the overall width of the distributions remains nearly constant at $\sigma_x=0.5$ m with no significant change with expulsion height. From wider dispersal fields using the LED lights as in Fig.~\ref{fig:1_schematic}(b), we find a similar distribution width $\sigma_x \approx 0.5$ m and also find that the maximal distribution along $y-$axis $\sigma_y \approx 0.6$\,m. 

To probe the nature of the distributions, we plot the same distributions in semi-log plot in the corresponding insets in Fig.~\ref{fig:3_distributions}(a-c). We find that $\phi$ decays exponentially after reaching their peak. The fact that the decay is exponential implies that the processes which disperse mucosaliva once they emerge as a cone and lose forward momentum are stochastic. 
Thus, we fit the distributions to the form 
\begin{equation}
f(x)=C  x^{\alpha} e^{-\beta x}\,,    
\label{eq:fx}
\end{equation}
where $C$, $\alpha$, and $\beta$ are fitting parameters listed in Table~\ref{tab:parameters}. It can be noted that this form captures the observed distributions very well, including the peak $\phi_m$ and mean $\bar{x}$ shown in Fig.~\ref{fig:3_distributions}(d) and Fig.~\ref{fig:3_distributions}(e), respectively. 

\begin{table}
\setlength{\tabcolsep}{10pt}
\begin{center}
\begin{tabular} {| c | c c c |}
\hline
$H$ (m) &  $C$  & $\alpha$ & $\beta$ (m$^{-1}$) \\
\hline
$0.5$ & 22    & 6   & 10    \\ [1ex]
$1.0$ & 9     & 5  & 9.5   \\ [1ex]
$1.5$ & 2.5   & 4 & 8.8    \\ [1ex]
\hline
\end{tabular}
\end{center}
\caption{Fitting parameter values for each distribution of the form shown in Eq.~(\ref{eq:fx}).}
\label{tab:parameters}
\end{table}

\begin{figure*}[ht!]
\begin{center}
\includegraphics[width=.8\linewidth, trim={0cm 12cm 7cm 0},clip]{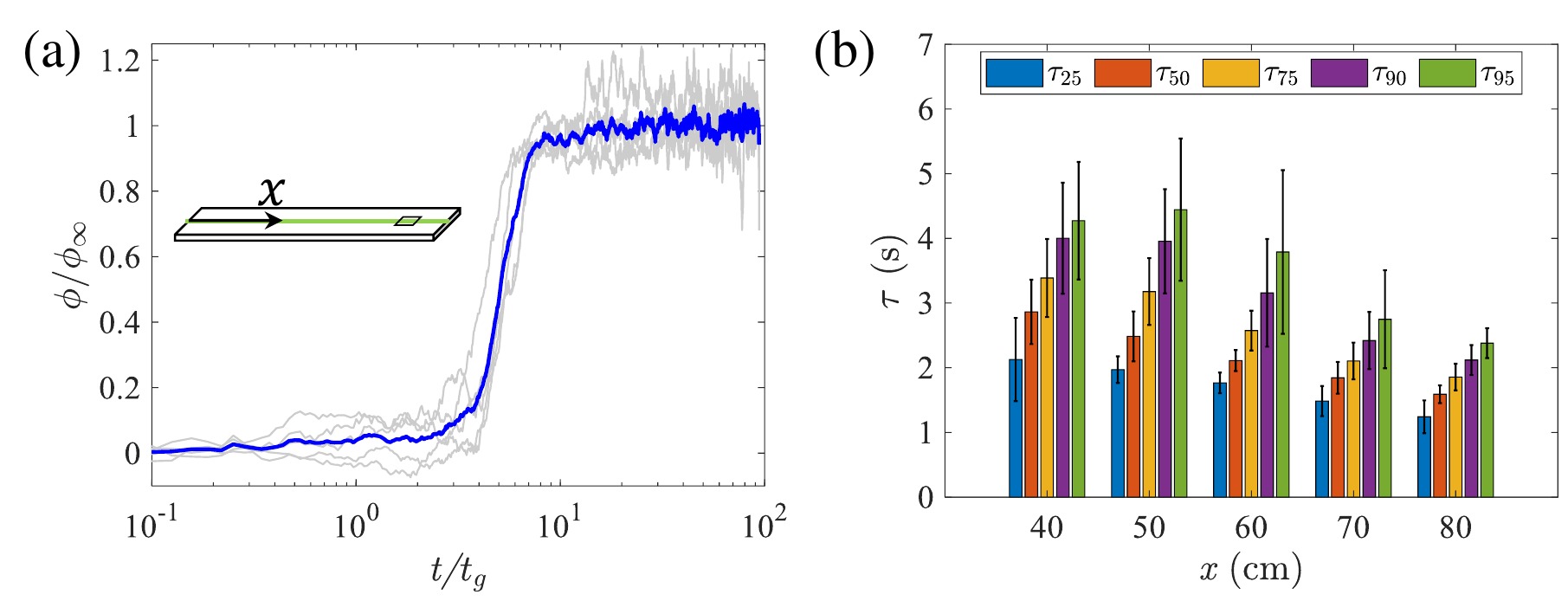}
\caption{ (a) The time evolution of the volume of droplets over a range of $x=[0.7, \, 0.8]$ m. Droplets are expelled from a height of $H=1.5$ m. Volumes are self-normalized by the maximum volume such that $\phi/\phi_\mathrm{\infty}$. Time is normalized by the gravitational settling time, $t_g=\sqrt{2H/g} \sim 0.55$\,s. Solid blue line is the average of $n_t=5$ trials (light gray lines). (b) The landing time, $\tau$, of droplets at $H=1.5$ \,m. The time it takes to reach a percentage of $\phi_\infty$, $\tau$, is plotted for $x=40-70$\, cm. At a given $x$, the time to reach 25\%, 50\%, 75\%, 90\%, and 95\% of $\phi_\infty$ goes from left to right. Error bars are one standard deviation from the mean.
}
\label{fig:4_landingTime}
\end{center}
\end{figure*}

We determine the distance $x^{(98\%)}$ along the $x$-axis where 98\% of the material has deposited and plot it both using the raw distributions and Eq.~\ref{eq:fx} in Fig.~\ref{fig:3_distributions}(f). Using both methods, we find that the distance where this point is reached is well within 2\,m, and decreases somewhat with increase in $H$. From Eq.~\ref{eq:fx}, we find that less than $0.0115\%$ of the mucosaliva launched launched from a height of 1.5 \,m directly in front of the face is expected to reach beyond $x = 2$\,m. This roughly equates to a volume of 86.6 \, $\mu$L, or a single droplet with a diameter of $14 \,\mu $m, which is close to our lower limit droplet size range.

It is further interesting to note that $\phi_m$, $\bar{x}$, and $x^{(98\%)}$ all trend higher with lower $H$.  This may indicate that the interaction of the droplet cloud with the substrate causes a  collective spread as in a gravity current~\cite{Benjamin1968}. It is also possible that falling from a greater height also leads to greater diffusion backwards as indicated by the streaks in the bottom-left corner of the image shown in Fig.~\ref{fig:1_schematic}(a). Further modelling is required to understand the origin of this surprising trend.

\subsection{Landing Times} \label{sec:landing}

To understand the contribution of the aerial history of the various droplets to the build up of the spatial distribution, we examine their landing times on the substrate. The normalized fraction $\phi/\phi_\mathrm{\infty}$ over a given distance is obtained as a function of normalized time, $t/t_g$. Here, $\phi_\mathrm{\infty}$ is the cumulative amount of mucosaliva reached over long times at the location of interest, and $t_g=\sqrt{2H/g}$ is the gravitational time scale. For $H=1.5$\,m, $t_g \approx 0.55$\,s.  Figure~\ref{fig:4_landingTime}(a) shows $\phi/\phi_\infty$ averaged between $x = 0.7$ and 0.8\,m as a function of time for a single spray ($n_s =1$) based on five trials corresponding to $H=1.5$\,m. Each trial is also shown in light gray to give a sense of the variation. We observe that $\phi/\phi_\infty$ starts to build up after $t/t_g \gtrsim 1$ indicating that almost all, except the largest droplets, are slowed down by air drag before they reach the floor. 
The increase in $\phi/\phi_\infty$ is rapid at first before becoming more gradual and continuing to increase slowly over time over several seconds. This can be seen in the SD Movie~3~\cite{sup-doc}. Over longer time scales, where $t/t_g \gtrsim 10$, we find that there is a small increase in the volume over time. This is likely due to the smallest droplets, i.e. aerosols, falling at a slower rate. However, these aerosols do not contribute significantly to the total volume after $t/t_g \sim 10$. 

To understand the relationship between the landing time distribution of the mucosaliva and the distance from source, we plot the time, $\tau$, taken for the droplets to accumulate from 25-95\% of its final value at each respective location in Fig.~\ref{fig:4_landingTime}(b). For example, at $x=0.4$\,m, the time it takes to accumulate 25\% of $\phi_\infty$ is $\tau_{25}=2.1 \pm 0.64$\,s and the time it takes to accumulate up to 95\% is $\tau_{95}=4.5 \pm 0.91$\,s. Thus, we can interpret $\tau_{25}$ as the very first droplets to land on the surface and $\tau_{95}$ as some of the last. We observe that landing times of the mucosaliva is systematically longer, closer to the source. In other words, the first droplets to land are doing so further away from the source. This may seem counter intuitive, but as was noted in the discussion of Fig.~\ref{fig:1_schematic}(a), the droplets that appear to travel in near parabolic paths, and thereby have negligible drag, travel further, whereas those which travel as a cloud tend to be smaller in size and are thus affected more greatly by air drag.

\section{Analysis of the settling time scales}
\subsection{Projectile dynamics}\label{sec:proj}
\begin{figure}
\begin{center}
\includegraphics[width=.85\linewidth, trim={0cm 0cm 21cm 0}]{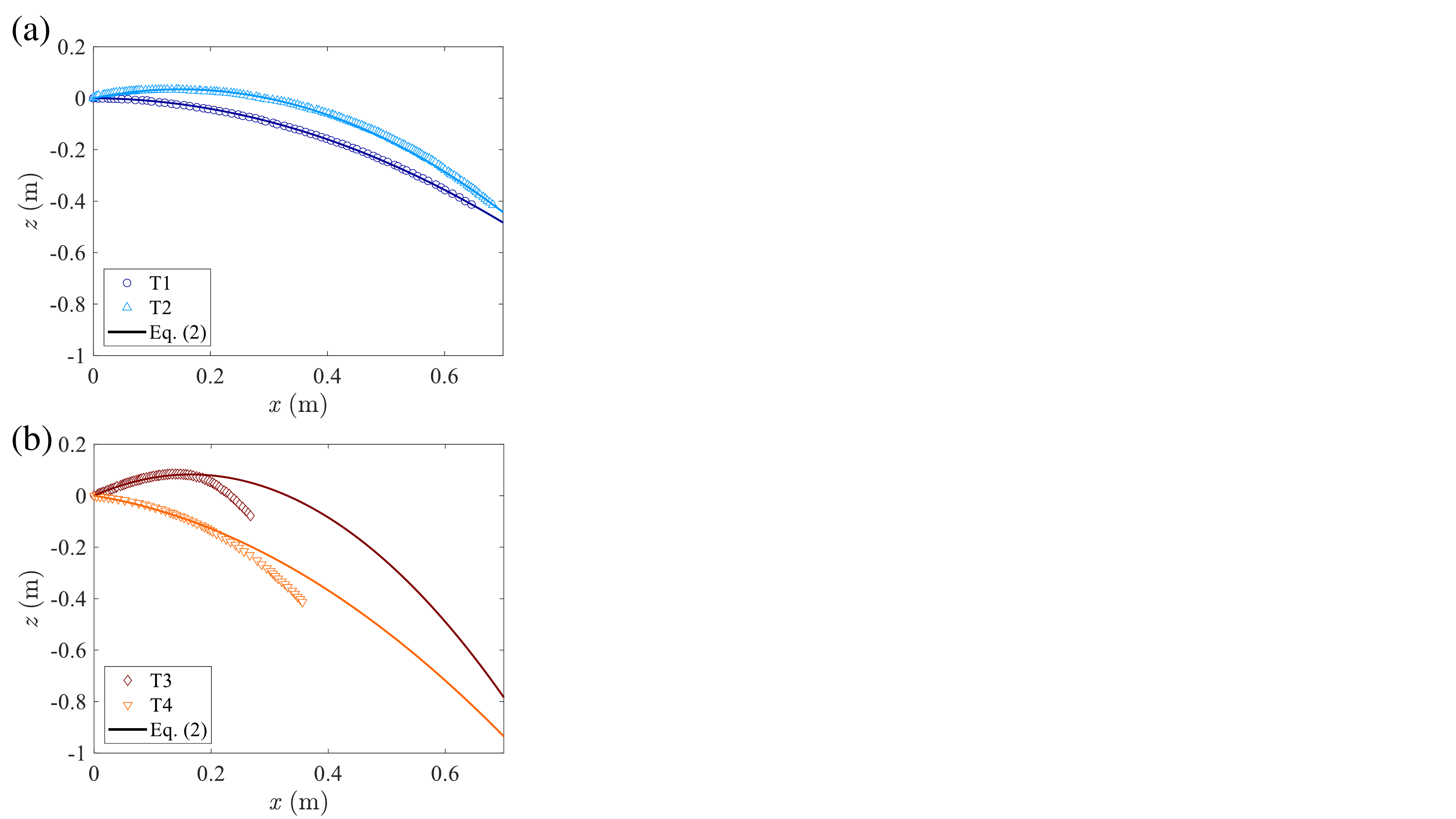}
\caption{{Trajectories T1-T4 of individual droplets corresponding to some of the brighter traces in Fig.~\ref{fig:1_schematic}(a). (a)~Some droplets travel with pure projectile motion as shown by good agreement with Eq.~(\ref{eq:proj}). (b) Trajectories T3 and T4 deviate from projectile motion  due to air drag. The initial conditions are for T1:~$\theta=-0.7^\circ$, $u_0=2.25$ m/s; T2:~$\theta=25^\circ$, $u_0=1.95$ m/s; T3:~$\theta=45^\circ$, $u_0=1.8$ m/s; and T4:~$\theta=-20^\circ$, $u_0=2.0$ m/s. }}
\label{fig:5_droptrajectories}
\end{center}
\end{figure}

If the drag force acting on a droplet is negligible, it follows simple projectile motion given by
\begin{equation}
z(x) = \tan(\theta) \, x - \frac{g}{2 u_0^2 \cos^2\theta} \, x^2,
\label{eq:proj}
\end{equation}
 where $\theta$ is the initial ejection angle, $u_0$ is the initial velocity, and $g$ is gravitational acceleration. Figure~\ref{fig:5_droptrajectories}(a) shows examples of tracked droplet trajectories which are well matched with Eq.~(\ref{eq:proj}) with no adjustable parameters. However, deviations from parabolic trajectories can be observed in examples shown in Fig.~\ref{fig:5_droptrajectories}(b).
 
 Because gravitational force scales as the cube of the radius, while drag scales as the square of the radius, gravity can be expected to dominate the trajectory of sufficiently large droplets over the observed range of speeds. Thus parabolic trajectories can be expected to correspond to the larger droplets, whereas air drag and momentum exchange with air can be expected to play an increasing role with decreasing droplet size. The good agreement depicted in Fig.~\ref{fig:5_droptrajectories}(a) indeed shows that at least some ejected drops show parabolic trajectories. However, the majority of the droplets appear as a cloud as it expands and settles slowly in gravity as was noted in the discussion of  Fig.~\ref{fig:2_cloudboundary}.

\subsection{Effect of drag}\label{sec:cloud}

To gain insight into conditions where air drag plays a prominent role, we consider the settling time of a homogeneous sphere with diameter $D_s$ and density $\rho_s$. As the sphere falls through air, the general form of terminal velocity of a sphere $u_s$ is found by balancing weight, buoyancy, and drag, and is given by,
\begin{equation}
u_s=\sqrt{ \frac{4}{3} \frac{g D_s}{C_d} \frac{\rho_s-\rho_a}{\rho_a} },
\label{eq:terminalvelocity}
\end{equation}
where $\rho_a=1.225$\, kg/m$^3$ is the air density and $C_d$ is the drag coefficient of a sphere as a function of its Reynolds number $Re$ given by {the form approximated from empirical formulations of the drag coefficient~\cite{yang2015general}}, 
\begin{equation}
    C_d(Re) = \frac{24}{Re} + \frac{4}{\sqrt{Re}} + 0.4\,,
    \label{eq:CdRe}
\end{equation}
valid for Reynolds number $Re<2\times10^5$ and approaches $C_d=24/Re$ for small $Re$.  Here,  
\begin{equation}
     Re = \frac{\rho_a D_s u_s}{\mu_a}. 
     \label{eq:Re}
 \end{equation}
 
 Due to the implicit nature of Eqn.~(\ref{eq:terminalvelocity}), (\ref{eq:CdRe}), and (\ref{eq:Re}), the terminal velocity is solved numerically.
 Considering a single droplet falling from a height of $H=1.5$\,m, a $D=10\,\mu$m droplet would fall at $u_s=0.003$\,m/s, which yields a settling time of $\tau=500$\,s. On the other hand, droplets on the order of $D=100-1000\, \mu m$ would fall at $u_s \approx 0.241$\,m/s to 3.78\,m/s, or landing times ranging from $\tau \approx 0.4$ to 6.2\,s. This range agrees well with what we observe in landing times in Fig. \ref{fig:4_landingTime}(b). But since individual size and {kinematics of each droplet} are unknown, we use the cloud boundary to estimate the falling droplets as a collective.

\subsection{Cloud dynamics}
\begin{figure}
\begin{center}
\includegraphics[width=.85\linewidth, trim={0cm 6.5cm 16cm 0cm},clip]{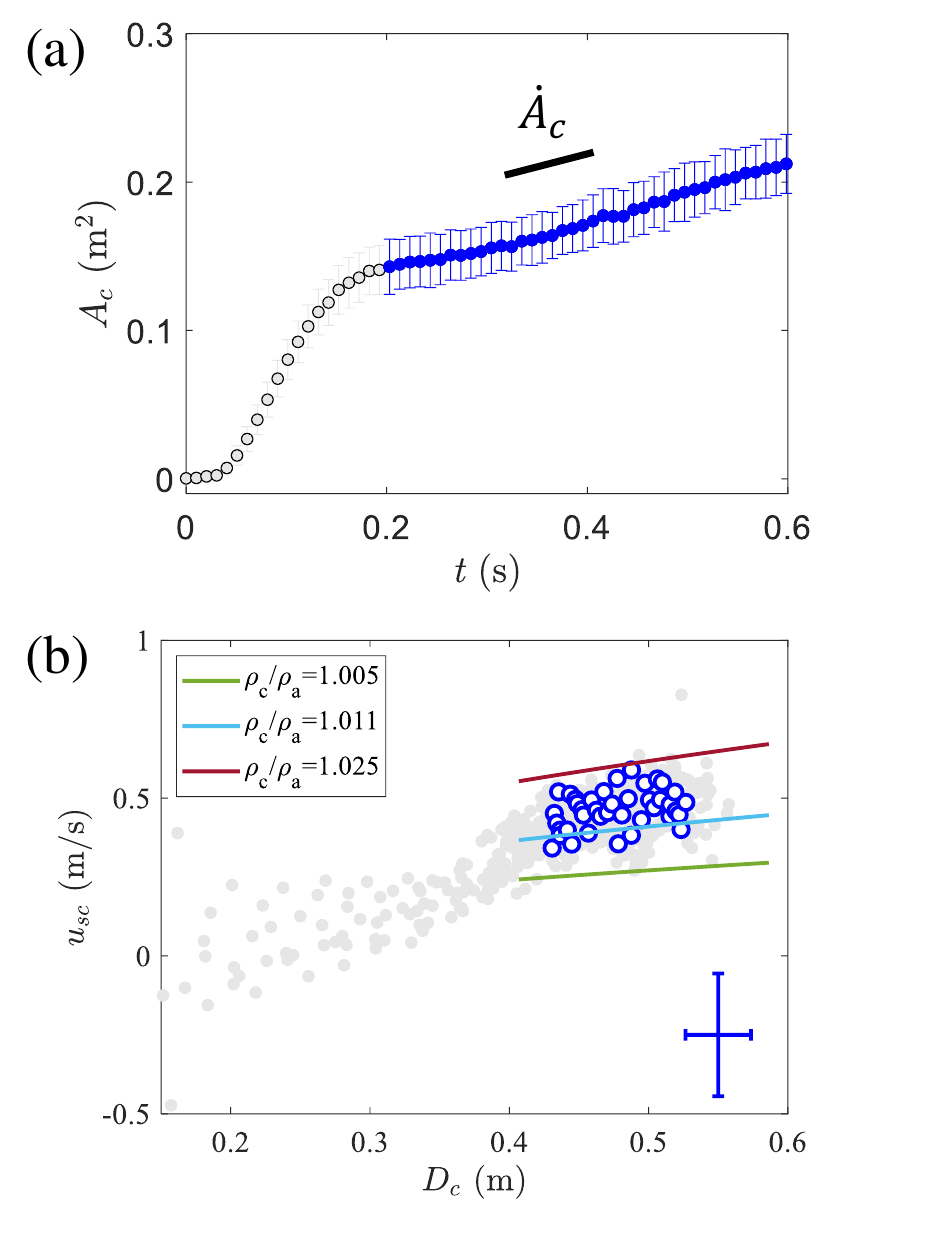}
\caption{(a) The average 2D projected area of the cloud, $A_c$, over time increases linearly after $t=0.2$\,s (blue circles) with a slope of $\dot{A}_c=0.2$\,m$^2$/s. Error bars are one standard deviation. (b) The average settling speed of the cloud relative to the average effective diameter, $D_c=\sqrt{4A_c/\pi}$ (blue circles) after $t>0.2$\,s. The lines are implicit calculations of the terminal velocity as a function of discrete diameters. Our data falls within the the calculations in which cloud density, $\rho_c$, is between 0.5-2.5$\%$ greater than the density of air, $\rho_a$. An estimated density ratio of $\rho_c/\rho_a$= 1.011 agrees well with the experimental data. Gray circles indicate the the cloud growth for all $n_t=10$ trials. One standard deviation for diameter and velocity is shown in lower right corner.}
\label{fig:6_cloudSettlingSpeed}
\end{center}
\end{figure}
While considering the cloud of droplets as a sphere, $\rho_s = \rho_c$, the cloud density, $D_s = D_c$, the effective cloud diameter, and $u_s = u_{sc}$, the cloud settling speed. Then, we estimate $D_c=\sqrt{4A_c/\pi}$, based on the 2D projection area, $A_c$, of the cloud as it evolves in time.  Fig.~\ref{fig:6_cloudSettlingSpeed}(a) shows  $A_c$ as a function of time. After the initial growth period corresponding to the duration of the sneeze ($t<0.2$ s), $A_c$ grows at a constant rate of 0.2 m$^2$/s, which is also comparable to what is found in human sneezes~\cite{Tang2013}. 

Once again, we numerically solve Eqn.~(\ref{eq:terminalvelocity}), (\ref{eq:CdRe}), and (\ref{eq:Re}) to calculate the cloud terminal velocity, $u_sc$. The terminal velocity converges for discrete values of $D_c$ after a few iterations. This set of calculations reveals a range of settling velocities for various cloud densities. As can be seen from Fig.~\ref{fig:6_cloudSettlingSpeed}(b),  our data mostly sits in between the lines where $\rho_c$ ranges between 0.5\% and 2.5\% of $\rho_a$ after the initial growth period. The density of the cloud can be also estimated by calculating the mass of air and the mass of a single spray, which yields an average density of $1.2387 \pm 0.0027$\,kg/m$^3$, or a density ratio of $\rho_c/\rho_a=1.011$. Based on these calculations, we find that the velocity of the cloud $u_{sc}$ is $0.38 \pm 0.078$\,m/s, which is in agreement with our experimental observations of cloud dynamics shown in Fig.~\ref{fig:2_cloudboundary}(d). Given an expulsion height of $H=1.5$\,m, the landing time would be in the range $\tau=3.3-4.9$ \,s, which is also within an order of magnitude of our observations on the landing times in Fig.~\ref{fig:4_landingTime}(b). 

Moreover, considering that the majority of droplets contained in the cloud lie between the geometric center $x\approx0.4$ and the boundary $x\approx0.7$\,m, then the majority of droplets as captured by considering  $\tau_{90}$ or $\tau_{95}$ land within the calculated landing times. This is a reasonable estimation, particularly given our assumption that the cloud is a homogeneous sphere. 

\section{Mask Efficacy} \label{sec:mask}
\begin{figure}
\begin{center}
\includegraphics[width=.8\linewidth, trim={0 1cm 16cm 0}]{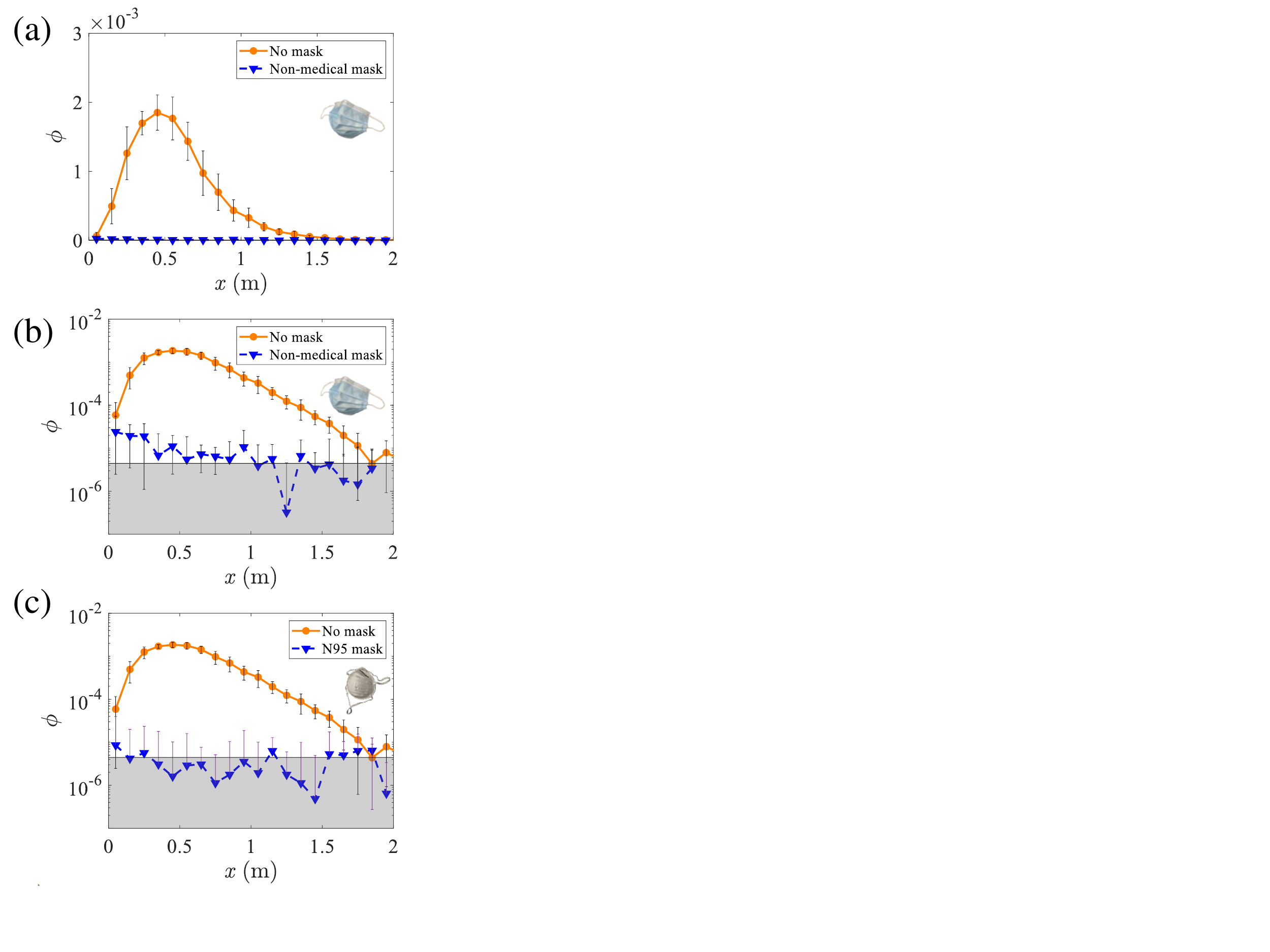}
\caption{(a) Comparing the distribution profile of mucosalivary droplets for no mask vs with mask  from a height of 1.5 m. (b) Same comparison as above on semilog scale. $n_t=5$ trials for both cases. {(c) Switching to a N95 mask reduces the dispersal to within the noise floor of our measurements. Wearing a standard non-medical or N95} mask significantly reduces the total volume by at least 99\%, and reduces the peaks recorded by at least 99.5\%. Noise floor is $\phi_{noise}=4.4\times10^{-6}$, as shown in gray.}
\label{fig:7_mask-or-nomask}
\end{center}
\end{figure}

We now examine the effectiveness of wearing a standard non-medical nose and mouth mask under repeated sneezing events {and a N95 mask} (see insets in Fig.~\ref{fig:7_mask-or-nomask}(a-c)) using the same system as were used in the discussion in the previous sections without masks. We visualize mucosalivary sprays emerging at a height $H = 1.5$\,m, using both high speed imaging as well as the deposition of the material on the ground in front of the source with fluorescence imaging. A movie comparing the droplets emerging from unmasked face and one covered with the mask can be found in the SD Movie~4~\cite{sup-doc}. No droplets are visible beyond the confines of the mask, in stark contrast to the case with no mask on. Similar observations have been made in the case of speech~\cite{Anfinrud2020} recently where the expiration speeds are smaller. However, this technique can miss vanishingly small droplets because they do not scatter sufficient light especially while doing high speed imaging.   

Therefore, we also examined the laser illuminated surface in front as a function of distance $x$. The data for the masked and unmasked case is plotted in linear and in semi-log scales in Fig.~\ref{fig:7_mask-or-nomask}(a-c).  
To find any trace amount of material, the results after trials consisting of 20 consecutive mucosalivary sprays ($n_s =20$) are examined in each case.  With the mask on, we find that trace amounts of droplets or aerosols could be identified above the noise threshold of $\phi_{noise}=4.4\times 10^{-6}$ close to the source below $x \approx 1$\,m. It is also possible that very small droplets or aerosols escape past the mask and stay aloft for long periods of time, spreading inside the lab space in which the experiments were conducted over time. However, no trace of these were detected even over days of experimentation. Clearly most of the sprayed mucosaliva was contained by the mask and stayed pooled inside after these multiple recorded events. 

By comparing maximum $\phi$ recorded with mask on with $\phi_m$ recorded in the case without the mask in Fig.~\ref{fig:7_mask-or-nomask}(b), we find that it is at least a factor 100 times lower. {The total volume of fluid that gets deposited on the laser line without a mask is approximately 8.84 $\, \mathrm{\mu}$L, whereas wearing a mask reduces that volume to 0.109$\, \mu$L for a standard non-medical mask and 0.0036 $\, \mu$L for an N95 mask.} Further, by integrating the measured $\phi$ from 0 to 2\,m in the case with the mask, and without the mask, we find that the ratio is less than 0.01. Thus, we conclude that at least 99\% of the mucosaliva which would be expelled forward is contained by wearing the mask.

\section{Conclusion} \label{sec:conclusion}
In summary, we have investigated mucosalivary droplet dispersal through the air and their deposition on a horizontal substrate using mechanically generated droplet sprays tuned to correspond to asymptomatic human adult sneezes. Two complementary imaging techniques were used to examine the dispersal of the droplets in air, and as they settled onto a horizontal substrate. By high speed imaging of the mucosalivary droplets using light scattering, we showed that the synthetic sneezes using bovine mucin and sprays from the mouth of a 3D printed mannequin face is consistent in terms of the droplet speed and sizes, the initial spray cone angle, and its expansion rate with published human sneeze examples~\cite{Bourouiba2014,Bahl2020b,Tang2013}. Leveraging the reproducibility and robustness of these mechanically generated mucosalivary sprays, we are able to (i) examine the dispersal of the droplets to a high degree of precision with statistical averaging performed over required number of trials, (ii) examine the effect of expiration height corresponding to sitting or standing, and (iii) quantify the effect of wearing a mask under reproducible conditions. Furthermore, the ability to add fluorescent dye to the synthetic mucosaliva as a passive tracer of a virus not only enables us to examine vanishingly small amounts of contamination deposited on surfaces, but also correct for any evaporation effects which can further render any small droplets invisible.     

We demonstrate that the aerial dispersal consists of a spectrum of droplets which at the largest sizes show essentially fast, parabolic projectile motion undisturbed by the air, and collective droplet cloud dynamics, which fall to the ground experiencing gravity, buoyancy, and drag as they fall through the ambient air over much longer time scales. We implement a simple numerical model to calculate the cloud terminal velocity, which matches well with our experimental results. Combined, we demonstrate that 95\% of the expelled material is deposited onto the surface within 5 seconds under still air conditions. It is possible that even slower settling dynamics corresponding to very small droplet or aerosols can play a further role in the remaining deposition dynamics.

By direct imaging, we find that the droplet cloud's geometric center travels up to 0.4 m, whereas the outermost boundary of the cloud travels to approximately 0.6 m. This differs from estimates reached by examining the deposited mucosaliva on horizontal surfaces which give higher mean dispersal distances. In particular, we find that spatial dispersals are broadly distributed with a width of approximately $0.5$\,m in the direction of the sneeze with a peak located at $x=$0.71\,m from a height of 0.5\,m, and $x=$0.57\,m from a height of 1.5\,m. Nonetheless we find that 0.0115\% of the expelled droplets, which is equivalent to a single 14 $\mu$m droplet, travels further than 2\,m, the typical distance given for mitigation of respiratory disease.

\begin{figure}
\begin{center}
\includegraphics[width=.8\linewidth, trim={0 0cm 15cm 0}]{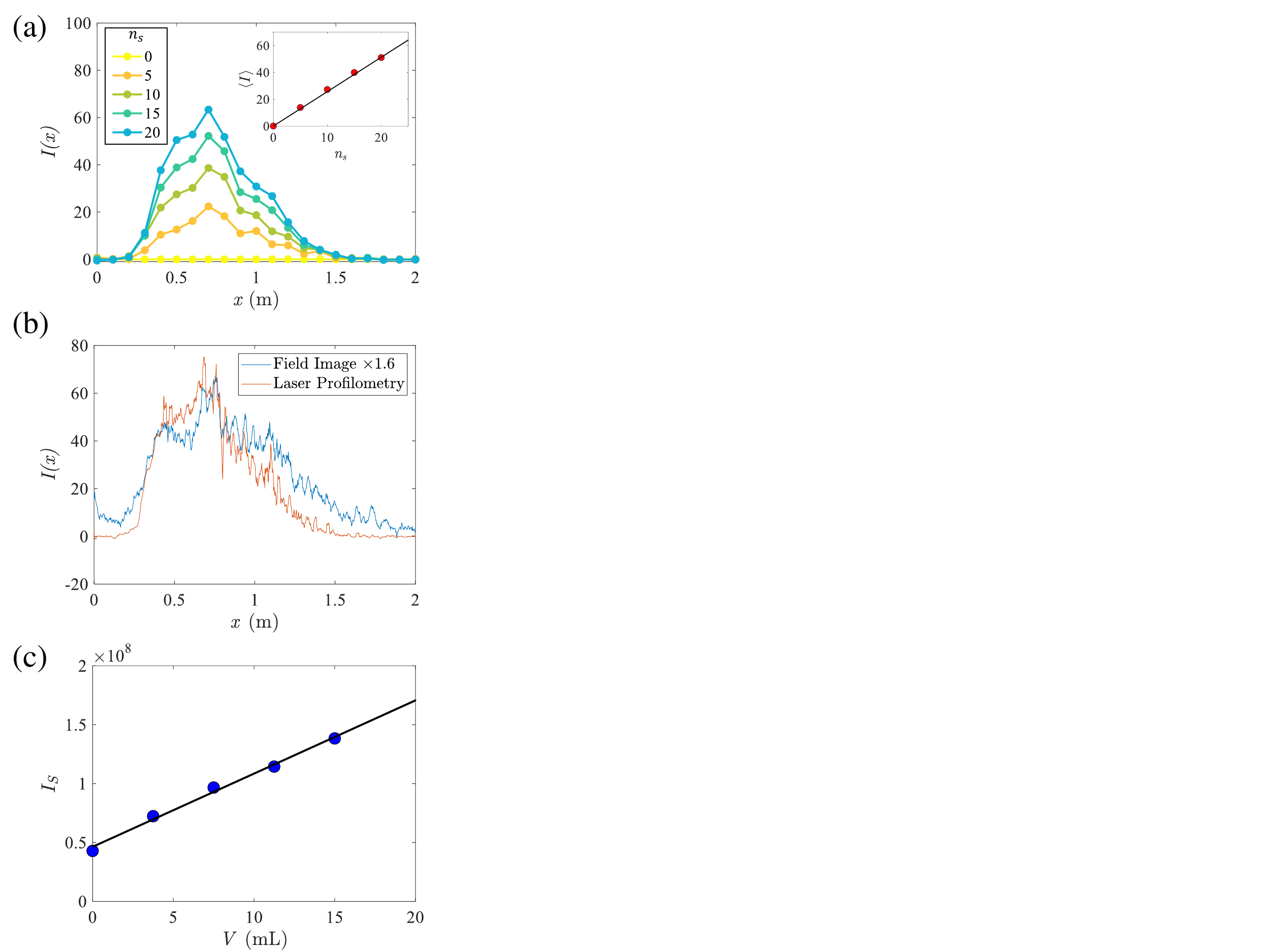}
\caption{Intensity calibration procedure. (a) The average intensity profile as a function of distance, $I(x)$ taken with laser profilometry after various $n_s$. Inset: The average intensity, $\langle I \rangle$, from $x=0.5$ to $x=0.6$ m, as a function of the number of sprays $n_s$, is well described by a linear fit with slope $2.56$.  (b) Matching the field image profilometry with laser profilometry. Field image profile is multiplied by a factor of 1.6 so that the maximum intensities are nearly equal. (c) The integrated intensity over the deposited area shown in Fig.~\ref{fig:1_schematic}(b). Integrated intensity and the volume of droplets have a linear relationship with slope $6.21 \times 10^6$. Volume per spray is $V_s=0.75$\,mL such that $V_{20} = n_s V_s \approx 15$\,mL. }
\label{fig:8_intensityCalibration}
\end{center}
\end{figure}

Finally, we have further demonstrated that using a standard nose and mouth mask reduces the dispersal of droplets by a factor of a hundred using highly sensitive fluorescence imaging mostly in the form of very small droplets which are invisible to the naked eye. The droplets are very small and only trace amounts of droplets are visible past the mask under otherwise similar conditions. { We demonstrate that while a N95 mask indeed reduces the volume of mucosaliva dispersed to within the noise floor of our measurements, even a well fitted mask does remarkably well in containing most of the volume of mucosaliva dispersed.}

This study is designed under still-air conditions and caution should be used in interpreting the results in enclosed spaces with HVAC where draft speeds of 0.1 to 0.2 m/s are considered normal for comfort. Given that a small fraction of the mucosalivary spray can stay aloft for five seconds or more increases the likelihood of its spread over far greater distances. Without dilution, this can lead to build up of small droplets or aerosols in the air over time leading to possible infection in the case of respiratory diseases~\cite{Lu2020}.

\appendix

\section{Imaging}\label{sec:appendix}
Various combinations of illumination and imaging techniques are used to visualize the dispersal and deposition of the mucosaliva. The spray is visualized from the side using a Phantom VEO-E 310 Monochrome camera, and light from high intensity LED arrays that scatter from the droplets. These images are used in measuring the dispersal of the droplets while they move through air. Complementary imaging methods are used to visualize the mucosaliva deposited on surfaces. 

A Pixelink Color PL-D7512 camera captures a 2\,m long region using a long exposure time of 5\,seconds to capture even trace amounts of mucosalivary droplets deposited on the surface. Further, a full field view of the deposited mucosaliva is imaged using uniform LED illumination, which gives a less sensitive measurement compared to laser lighting, but gives the overall shape of the dispersal lobes observed after the spray settles onto the horizontal surface. Taking advantage of the observed linearity of light intensity with spray volume shown in Fig.~\ref{fig:8_intensityCalibration}(c), we are able to obtain very sensitive measurements of the deposited mucosaliva which is not possible by direct visualization of the sprays themselves.

Fig.~\ref{fig:8_intensityCalibration}(a) shows the average intensity $\langle I \rangle$ recorded at distances between $x = 0.5$\,m and $x=0.6$\,m as a function of sprays. The sprays were released at least one second apart to allow the droplets to fall to the ground without interference.  The data is well described by a linear fit, which is then used to calibrate the intensity to the amount of medium deposited at a given location on the surface. We compare the intensity profiles after 20 sprays using an LED array and a laser sheet in Fig. \ref{fig:8_intensityCalibration}(b). By multiplying the LED intensity profile by a factor of 1.6, we show that the peak intensity matches well. The integrated intensity, $I_S=\int_S I(x,y) dA$, over the area of the $x-y$ plane then enables us to calculate the relationship between intensity and volume of droplets per unit area.

\begin{acknowledgments}   
The authors would like to thank Emily Chang, Sujata Davis, Anthony Gai, and Alexander Petroff for their helpful comments. This work was supported by National Science Foundation COVID-19 RAPID Grant No. DMR-2030307. 
\end{acknowledgments}   



\end{document}